\documentclass{llncs}
\usepackage{IEEEtrantools}

\usepackage{paralist}
\usepackage[inline]{enumitem}
\usepackage{textcomp}

\usepackage{hyperref}

\usepackage[dvipsnames]{xcolor}

\usepackage{tikz}
\usetikzlibrary{arrows,calc}

\usepackage{amsmath}
\usepackage{amssymb}

\usepackage{nicefrac}
\usepackage{booktabs}
\usepackage{multirow}

\usepackage{macros}

\title{Quantitative Implementation Strategies\\ for Safety Controllers\thanks{\
Partially supported by the DFG project ``Game-based Synthesis for Industrial Automation'' and
the Institute for Advanced Study of the Technische Universit{\"a}t M{\"u}nchen.
}}
\titlerunning{}  
\author{%
  Philipp J. Meyer \inst{2} \and 
  Matthias Rungger\inst{1} \and 
  Michael Luttenberger\inst{2} \and \\
  Javier Esparza\inst{2} \and
  Majid Zamani\inst{1} 
}
\authorrunning{P.\,J.~Meyer et al.} 
\institute{Professorship for Hybrid Control Systems\\
Department of Electrical and Computer Engineering\\
Technical University of Munich\\
\and
Chair for Foundations of Software Reliability and Theoretical Computer
Science\\
Department of Informatics\\
Technical University of Munich
}

\begin{document}
\frontmatter          
\pagestyle{headings}  
\mainmatter              
\maketitle              

\begin{abstract}

We consider the symbolic controller synthesis approach to enforce safety
specifications on perturbed, nonlinear control systems. In general, in each
state of the system several control values might be applicable to enforce the
safety requirement and in the implementation one has the burden of
picking a particular control value out of possibly many.
We present a class of implementation strategies to obtain a controller
with certain performance guarantees.
This class includes two existing implementation strategies from the literature,
based on discounted payoff and mean-payoff games.
We unify both approaches by using games characterized by a single discount factor
determining the implementation.
We evaluate different implementations from our class experimentally on two case studies.
We show that the choice of the discount factor has
a significant influence on the average long-term costs,
and the best performance guarantee for the symbolic model does not result in the
best implementation.
Comparing the optimal choice of the discount factor here with the previously proposed
values, the costs differ by a factor of up to 50.
Our approach therefore yields a method to choose systematically a good implementation for safety
controllers with quantitative objectives.

\end{abstract}

\section{Introduction}
\label{s:introduction}

The symbolic controller synthesis approach~\cite{Tabuada09} has gained considerable
attention within the Cyber-Physical System research community. The approach
constructs a finite-state abstraction (a.k.a. \emph{symbolic model}) of a
continuous-space, continuous-time system,
and reduces the control problem over the continuous system to the computation of a
winning strategy in a finite-state game. Following this paradigm, it is guaranteed by construction that the winning
strategy delivers a correct controller for the concrete system, i.e., the refined controller enforces the
given specification over the original system.

In this work, we follow the symbolic approach to enforce safety specifications
for perturbed, nonlinear control systems. Safety specifications are one of the
most fundamental requirements and are ubiquitous in the analysis and control
of dynamical systems~\cite{Aubin91,BlanchiniMiani08}.
Given a system, the main goal is to synthesize a controller that enforces the
system to evolve within a safe set for all times.
Via the symbolic approach, the possible behaviours of the continuous
system
are safely over-approximated leading to a 2-player safety game over a finite (but large) arena.
The controller plays against an adversary that represents the nondeterminism,
which arises from the discretization of the
continuous system and possibly from model uncertainties and perturbations.
From a theoretical perspective, solving safety games over finite arenas is trivial: one simply
has to identify and then (iteratively) deactivate all actions which allow the
adversary to force a play out of the safety region. 
After deactivating all {\em dangerous} actions, the controller is essentially free
to play in any way within the winning region. It is well-known that there exists a unique {\em maximal}
(or {\em permissive} or {\em nondeterministic}) optimal winning strategy, which
assigns to each state all possible actions 
available to the controller that
guarantee that the system stays within the safety region. In general, there might
be multiple actions available to the controller and
the natural question is whether
we can make use of this flexibility
to further optimize the controller with respect to some quantitative objectives.
Examples of such quantitative conditions that arise quite naturally in the control of physical systems are:
\begin{itemize}
\item minimization of the number of input switches (lazy controller);
\item minimization of the deviation from a reference value (reference tracking);
\item minimization of energy consumption;
\item minimization of relative change between inputs (avoiding jerkiness).
\end{itemize}
%
%
In this paper, we follow ideas in
\cite{RunggerReissigZamani16,MeyerGirardWitrant15} and use possibly discounted
mean-payoff games to synthesize deterministic controllers that enforce the
system to stay within the safety region and minimize certain cost functions.
Specifically, we consider  {\em normalized $\lambda$-discounted cost functions}
(with $\lambda\in[0,1]$)
\begin{equation}\label{e:normdisc}
(1-\lambda) \sum_{i=0}^\infty \lambda^i c(i)
\end{equation}
and {\em limit-average cost functions}
\begin{equation}\label{e:limavg}
\limsup_{T\to \infty} \frac{1}{1+T}  \sum_{i=0}^T c(i)
\end{equation}
where $c(i)$ represents a certain cost (to be defined later) that is associated with the system
behavior at time $i$.
We follow standard nomenclature, and call the resulting
games whose goal is to minimize those cost functions {\em $\lambda$-discounted
payoff safety games ($\lambda$-DPSG)} and {\em mean-payoff safety game (MPSG)}, respectively.

We set up an effective tool chain for symbolic controller synthesis to enforce
safety specifications under mean-payoff objectives.
In the first step, we employ SCOTS~\cite{RunggerZamani16}
to obtain the symbolic model; subsequently, we compute the winning strategies for the resulting DPSGs and MPSGs
with the state-of-the-art GPU solver described in~\cite{MeyerLuttenberger16}; finally,
the performance of the controller in the continuous-time system is evaluated by means of simulations
conducted with MATLAB. We measure the performance by the limit-average cost~(\ref{e:limavg}) for
controllers obtained from both DPSGs and MPSGs.

We present two case studies and conduct a systematic analysis to
illustrate the effects of different choices of cost functions and
discount parameters. In the first case study, we consider the regulation of the room temperature
and the humidity of the air in a building equipped with a
heating, ventilation and air conditioning
(HVAC) system introduced
in~\cite{BrocchiniFalsoneManganiniHolubPrandini16,HolubZamaniAbate16}. In the
second example, we consider regulation of the popular  cart-pole system, e.g. studied in~\cite{Reissig10}.
For each of the different objectives, we associate both $\lambda$-discounted cost functions and limit-average cost functions,
and solve the corresponding games.
Our main finding is that
\emph{performance guarantees come at a price}:
while the limit-average cost function provides the strongest,
global performance \emph{guarantee} on the actual limit-average cost over the complete execution of the controller, it typically leads to controllers
that perform poorly. Using discounted cost functions leads to controllers with only
weaker limit-average cost guarantees, which however achieve
better limit-average costs in the simulation. We also present experimental results about the values of $\lambda$ leading to
controllers with the best limit-average costs.

\paragraph{Related work.} Quantitative objectives are a natural requirement in specifying the desired
system behavior. Consequently, there exist several different approaches to
augment language containment specifications, e.g.\ defined in linear temporal
logic, with quantitative
properties~\cite{Girard12,GolLazarBelta15,MazoTabuada11,MeyerGirardWitrant15,ReissigRungger13,TazakiImura12,WolffTopcuMurray12,WolffMurray16}.
A considerable number of approaches focus on finite horizon specifications, like
reachability~\cite{Girard12,MazoTabuada11,ReissigRungger13,TazakiImura12}
or co-safety~\cite{GolLazarBelta15} specifications, which leads to an optimization of
the transient behavior. Our work, on the contrary,  focuses on optimizing
the average, long-term behavior. This is also the focus of~\cite{WolffTopcuMurray12,WolffMurray16}.
However, the approach of these papers is incomparable
with ours: while their work is applicable not only to safety
specifications, but also to more general temporal logic specifications, it is
restricted to specific classes of systems. In particular, \cite{WolffTopcuMurray12} is
restricted to deterministic symbolic models, and \cite{WolffMurray16} is restricted to mixed
logical dynamical systems and differentially flat systems. None of these classes contains
the control systems considered in this work.
While DPSG and MPSG in the context of symbolic synthesis have been considered in
\cite{MeyerGirardWitrant15} and \cite{RunggerReissigZamani16}, this work
is the first one to systematically study the effects of different cost functions
and discount factors on the performance of the resulting refined controllers.

\paragraph{Structure of the paper.}
Section~\ref{s:control-problems} presents the control problem.
Section~\ref{s:symbolic-synthesis} presents the symbolic approach to controller synthesis.
Section~\ref{s:games} presents the game-theoretic approach to quantitative safety synthesis.
Section~\ref{s:case-studies} describes the cost functions and our two case studies: the HVAC system and the cart-pole system.
Section~\ref{s:experiments} presents the results of our experimental evaluation.
Section~\ref{s:summary} summarizes our findings.

\section{Control Systems and Objectives}
\label{s:control-problems}
We study nonlinear control systems of the form
\begin{IEEEeqnarray}{c}\label{e:sys:ct}
\dot \xi(t) \in f(\xi(t),u) + \segcc{-w,w}
\end{IEEEeqnarray}
where $f$ is given by \mbox{$f:\mathbb{R}^n\times \bar U\to \mathbb{R}^n$} and
$\bar U\subseteq \R^m$. The vector $w=\intcc{w_1,\ldots,w_n}\in \mathbb{R}_{\ge0}^n$ is a perturbation
bound and denotes the hyper-interval
\begin{equation*}
 \segcc{-w,w}:=\intcc{-w_1,w_1}\times\ldots\times \intcc{-w_n,w_n}. 
\end{equation*}
We can use the perturbations to take into account model uncertainties and other adversarial
effects.
Given an interval $I\subseteq \R$, we define a \emph{solution}
of~\eqref{e:sys:ct} on $I$ under (constant) input
\mbox{$u\in \bar U$}
as an absolutely continuous function \mbox{$\xi \colon I \to \mathbb{R}^n$} that satisfies
\eqref{e:sys:ct} for almost every (a.e.) \mbox{$t \in I$}~\cite{Filippov88}.

\begin{figure}[h]
\centering
\begin{tikzpicture}
[
mynode/.style={draw,
               thick,
               rounded corners,
               fill=white,
               inner sep=.3cm,
               minimum width=1.5cm},
mynode1/.style={draw,
               semithick,
               rounded corners,
               inner sep=.3cm},
to/.style={->,
           >=stealth',
           shorten >=1pt,
           semithick,
           font=\sffamily\footnotesize},
]

\node [mynode] (sys) at (0,0) {$\dot \xi=f(\xi,\nu)+\segcc{-w,w}$};
\node [mynode] (con) at (0,-2.5) {Controller};

\node [mynode1] (zoh) at (-2.5,-1) {$\mathsf{ZOH}$};

\node [mynode1] (sam) at (2.5,-1) {\phantom{ZOH}};
\node at (2.25,-1) {$\tau$};

\draw[rounded corners,dashed,thick] (-3.75,-1.7) rectangle (3.75,.75);
\node at (0,-1.5) {$S$};

\draw [to,->]  (zoh.north) |-  (sys.west);
\draw [semithick]  (sys.east) -| (2.5,-.75);

\draw [to,->]  (con.west) -|  (zoh.south);

\draw[semithick] (2.5,-.75) -- (2.75,-1.15);
\draw[dashed,bend left,->] (2.8,-.8) -- (2.5,-1.15);

\draw[semithick,to] (2.5,-1.25) |- (con.east);

\end{tikzpicture}
\caption{Sample-and-hold implementation of a controller.}\label{f:sampleandhold}
\end{figure}
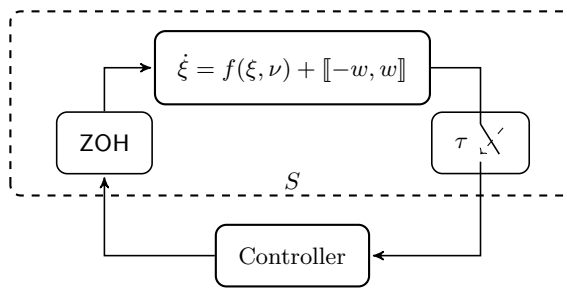

In order to facilitate a digital controller implementation, which operates in discrete time,
we consider the sampled behavior of the continuous-time system~\eqref{e:sys:ct}.
Let $\tau\in\R_{>0}$ be the sampling time. We assume that the
controller has access to the system state only at integer multiples of the
sampling times $k\tau$, $k\in\Z_{\ge0}$. Similarly, we assume
that the control input is updated only at the sampling times $k\tau$, $k\in\Z_{\ge0}$
and held constant otherwise. See Figure~\ref{f:sampleandhold} for an
illustration of this process. We cast the sampled behavior of~\eqref{e:sys:ct}
as a simple system \cite{ReissigWeberRungger16}.

For sets $A$ and $B$, a \emph{set-valued map} from $A$ to $B$ is a set $f \subseteq A \times B$,
written as $f : A \rightrightarrows B$, with $f(a) := \{ b \in B \mid (a,b) \in f \}$
for every $a \in A$.

\begin{definition}
A \emph{simple system} (with initial states) $S$ is a quadruple
\begin{IEEEeqnarray}{c}\label{e:sys}
  S:=(X,X_0,U,F)
\end{IEEEeqnarray}
where
the \emph{state alphabet} $X$, the \emph{initial state alphabet}
$X_0\subseteq X$, and the \emph{input alphabet} $U$ are nonempty sets, and 
the \emph{transition function} $F$ is a set-valued map  $F:X \times
U\rightrightarrows X$.
The \emph{behavior} of the simple system $S$ is defined by
\begin{multline*}
  \behavior{S}:=
  \big\{
    (u,x)\in (U\times X)^{\intco{0;T}} \mid x(0)\in X_0,\;\\
    \forall_{t\in\intco{0;T-1}}\; x(t+1)\in F(x(t),u(t))\\
    \text{ and }  
    (T<\infty \implies F(x(T-1),u(T-1))=\emptyset)
  \big\}.
\end{multline*}

\end{definition}

We say that a simple system $S$ of the form~\eqref{e:sys} \emph{represents the
$\tau$-sampled behavior of~\eqref{e:sys:ct}}, if $X=\R^n$, $U=\bar U$ and the transition function
satisfies for all arguments 
\begin{equation}
  F(x,u)=\{x'\mid \exists{\text{ a solution $\xi$ of~\eqref{e:sys:ct} on
  $\intcc{0,\tau}$ under $u$}}: \xi(0)=x \land \xi(\tau)=x'\}.
\end{equation}

A \emph{safety specification} for~\eqref{e:sys} is
simply a set  $Z\subseteq U\times X$.
A simple system $S$ together with a specification $Z$ constitute a
\emph{safety problem} $(S,Z)$. 

Even though, in this framework, we obtain safety
guarantees only with respect to the $\tau$-sampled behavior of~\eqref{e:sys:ct},
it is straightforward to work with a slightly modified safety specification
in the synthesis procedure, which then implies safety guarantees with respect to
the continuous-time behavior of~\eqref{e:sys:ct}, see
e.g.~\cite{MaidensKaynamaMitchellOishi13}.

A \emph{solution} of a safety problem $(S,Z)$, where $S$ is given
in~\eqref{e:sys}, is a set-valued map $K:X\rightrightarrows U$ (i.e.\ $\forall x\in X\colon K(x)\subseteq U$) such that the
\emph{closed loop} $K\times S$, which is the simple system $(X,X_0,U,F_K)$, whose transition function is defined by
\begin{equation*}
  F_K(x,u):=\{x'\in X\mid x'\in F(x,u)\land u\in K(x)\}
\end{equation*}
\emph{satisfies} the safety specification, i.e.,
\begin{IEEEeqnarray*}{c}
  \behavior{K\times S}\subseteq Z^{\intco{0;\infty}}.
\end{IEEEeqnarray*}
In this context, we refer to $K$ as a \emph{controller} for $S$.
The \emph{domain} of a controller $K$ is the set $\domain{K} :=
\{ x \in X \mid K(x) \neq \emptyset\}$.

Let $K$ be a solution of a safety problem $(S,Z)$ with $S$ given in~\eqref{e:sys}.
An \emph{implementation} of $S$ and $K$
is a function $K_{\rm imp}: X \times U \to U$ that satisfies $K_{\rm
imp}(x,u)\in K(x)$ for all $x \in \domain{K}$ and $u\in U$.
The closed loop $K_{\rm imp}\otimes S$
resulting from the implementation $K_{\rm imp}$ is the simple system 
$K_{\rm imp}\otimes S:=(X\times U,X_0\times U,\{0\},F_{\rm imp})$, whose transition function is defined by
\begin{equation*}
  F_{\rm imp}((x,u),0):=\{(x',u')\in X\times U\mid x'\in F(x,u')\land u'= K_{\rm imp}(x,u)\}.
\end{equation*}
For the remainder, we identify the behavior of $K_{\rm
imp}\otimes S$ with the set that results
from the projection of $\behavior{K_{\rm imp}\otimes S}$ onto $(X\times
U)^{\intco{0;\infty}}$. Given this identification, it is straightforward to see
that the inclusion 
\begin{equation*}
\behavior{K_{\rm imp}\otimes S}
\subseteq 
\behavior{K\times S}
\end{equation*}
holds. Hence, any closed loop $K_{\rm imp}\otimes S$ resulting from an
implementation $K_{\rm imp}$ of $K$ satisfies the
safety specification $Z$.

Subsequently, we assume we are given a \emph{cost function} 
\begin{equation}\label{e:costs}
  c:U \times X \times U \to  \R
\end{equation}
and a discounting factor $\lambda \in \intcc{0,1}$
which we use to determine the implementation strategy. Specifically, we would like
to find an implementation $K_{\rm imp}$ of a solution $K$ of $(S,Z)$ that
minimizes the worst-case, discounted cost function 
\begin{equation}\label{e:valuedpg}
  J(K{_{\rm imp}}\otimes S,c,\lambda):=\sup_{(u,x)\in\behavior{ K_{\rm imp}\otimes S}}
    (1-\lambda) \sum_{i=0}^\infty \lambda^i c(u(t), x(t), u(t+1)).
\end{equation}
For the case $\lambda = 1$, we consider the above function with the limit value for
$\lambda \rightarrow 1^-$,
where it is converges (see e.g.~Appendix~H of~\cite{FilarVrieze96} for a proof) to the limit-average cost
given by
\begin{equation}\label{e:valuempg}
  J(K{_{\rm imp}}\otimes S,c,1):=\sup_{(u,x)\in\behavior{ K_{\rm imp}\otimes S}}
    \limsup_{T\to\infty} \frac{1}{1+T} \sum_{i=0}^T c(u(t), x(t), u(t+1)).
\end{equation}
The quadruple $(S,Z,c,\lambda)$ constitutes a \emph{valuated safety problem}.


\section{Symbolic Synthesis}
\label{s:symbolic-synthesis}

Within the symbolic controller synthesis
paradigm~\cite{ReissigWeberRungger16,Tabuada09,ZamaniPolaMazoTabuada17} a controller $K$ for $S$ is
not computed directly, but a finite representation  $\hat S$ of $S$ is used as
a substitute in the synthesis process.
The procedure is roughly summarized in
three major steps: first, a finite representation $\hat S$, i.e., the symbolic
model, of $S$ is computed; second, the synthesis problem is algorithmically solved
with respect to $\hat S$; third, the obtained solution $\hat K$ is \emph{refined} or
\emph{transferred} to a controller $K$ for $S$. 
The correctness  of this approach
is usually ensured by relating the plant $S$ with the symbolic model $\hat S$ by
a system relation. 
In this work we follow~\cite{ReissigWeberRungger16}, in which symbolic models are
related with the plant via feedback refinement relations. Notably feedback refinement
relations are appealing  as they facilitate a particularly easy controller
refinement procedure compared to other system
relations~\cite{ReissigWeberRungger16}, see also the controller refinement
equation~\eqref{e:refinement}.

Let $S$ be given in~\eqref{e:sys}.
Consider a simple system 
\begin{equation}\label{e:abs}
  \hat S:=(\hat X,\hat X_0, \hat U,\hat F)
\end{equation}
whose states $\hat x\in \hat X$, also referred to as
\emph{cells}, are subsets of $X$, i.e.,  $\hat x\subseteq X$, and whose input
alphabet $\hat U$ is a subset of $U$. The system $\hat S$ is a \emph{symbolic
model} of $S$, if there exists a strict\footnote{A relation $R\subseteq A\times
B$ is strict, if for every $a\in A$ there exists $b\in B$ so that $(a,b)\in R$.} relation $Q\subseteq X\times \hat X$ so that
for all $u\in \hat U$ and $(x,\hat x)\in Q$ we have
\begin{enumerate}
\item  $x\in X_0$ implies $\hat x\in \hat X_0$;
\item  $\hat F(\hat x,u)\neq\emptyset$ implies
$F(x,u)\neq\emptyset$ and $Q(F(x,u))\subseteq\hat F(\hat x,u)$.
\end{enumerate}
The relation $Q$ is called a \emph{feedback refinement relation} from $S$ to
$\hat S$. The first condition ensures that every cell $\hat x$ that is related
to an initial state of $S$ is an initial state of $\hat S$. The second condition
ensures, that if a state-input pair $(\hat x,u)$ is non-blocking, i.e., $\hat
F(\hat x,u)\neq \emptyset$, then  $(x,u)$ is non-blocking for every related
state $x$. Additionally, every cell $\hat x'$ that is related to a successor state
$x'\in F(x,u)$ is also a successor of $(\hat x,u)$, i.e., $\hat x'\in \hat
F(\hat x,u)$.

A safety specification $\hat Z$ for~\eqref{e:abs} is derived from a safety
specification $Z$ for~\eqref{e:sys} by
\begin{equation}\label{e:abs:spec}
\hat Z:=\{(u,\hat x)\in \hat U\times \hat X\mid \{u\}\times \hat x\subseteq Z\}.
\end{equation}
In the refinement of a controller for the symbolic model $\hat S$ to a
controller for the concrete system $S$, the
feedback refinement relation $Q$ is interpreted as a set-valued map
$Q:X\rightrightarrows \hat X$ and is used to translate concrete states $x\in
X$ to related abstract states $\hat x \in Q(x)$. The refined controller $K$ for
$S$ is then simply given by the composition of the map $Q$ with the controller
$\hat K:\hat X\rightrightarrows \hat U$ for $\hat S$ by $K:=\hat K\circ Q$. In
this context, $K$ is referred to as \emph{refined} controller. 
The
correctness of this procedure is ensured by the following result recalled from \cite[Thm.~VI.3]{ReissigWeberRungger16}.

\begin{theorem}
Let $(S,Z)$ be a safety problem
with $S$ given in~\eqref{e:sys}. Let $\hat S$ and $\hat Z$ be given according
to~\eqref{e:abs} and~\eqref{e:abs:spec}, respectively. Suppose that $Q$ is a
feedback refinement relation from $S$ to $\hat S$. If $\hat K$ solves $(\hat S,\hat
Z)$ then $K:=\hat K\circ Q$ is a controller for $S$ which solves $(S,Z)$.
\end{theorem}

Suppose that $\hat
K_{\rm imp}$ is an implementation of $\hat S$ and $\hat K$, then we obtain an
implementation of $S$ and $K$ by 
\begin{equation}\label{e:refinement}
 K_{\rm imp}(u,x):=\hat K_{\rm imp}(u,P(x))
\end{equation}
where $P:X\to \hat X$ picks for every $x$ a related cell $\hat x$, i.e.,
$(x,P(x))\in Q$ for all $x\in X$. Again, $K_{\rm imp}$ represents a refinement
of the implementation  $\hat K_{\rm imp}$ derived for the symbolic model $\hat S$ and
controller $\hat K$.

Let $(S,Z,c,\lambda)$ be a valued safety problem with $S$ given in~\eqref{e:sys}.
Consider the valuated safety problem $(\hat S,\hat Z,\hat c, \lambda)$ with $\hat S$ and $\hat
Z$ given according to~\eqref{e:abs} and~\eqref{e:abs:spec}, respectively. Suppose that $Q$ is a
feedback refinement relation from $S$ to $\hat S$ and 
\begin{equation}\label{e:cost-abstraction}
  c(u,x,u')\le \hat c(u,\hat x,u') 
\end{equation}
holds for all $(x,\hat x)\in Q$ and $u,u' \in \hat U$.
For $\lambda = 1$ in~\cite{RunggerReissigZamani16}
and for $\lambda \in \intco{0,1}$ in~\cite{MeyerGirardWitrant15},
it is shown that the
worst-case costs associated with the controller $K_{\rm imp}$  derived
in~\eqref{e:refinement} from an
implementation $\hat K_{\rm imp}$ of $\hat S$ and $\hat K$ (for any qualified function
$P$) is upper bounded by the worst-case costs associated with $\hat K_{\rm
imp}$, i.e., the following inequality holds:
\begin{equation}
J(K{_{\rm imp}}\otimes S,c,\lambda)\le J(\hat K{_{\rm imp}}\otimes \hat S,\hat c,\lambda)
\end{equation}

In this work, we use {\tt SCOTS}~\cite{RunggerZamani16} to compute symbolic models
$\hat S$ of systems $S$ that represent the $\tau$-sampled behavior of
continuous-time control systems~\eqref{e:sys:ct}. The feedback refinement
relation $Q$ is given by the set-membership relation, i.e., $(x,\hat x)\in Q$
iff $x\in \hat x$.


\section{Quantitative Safety Synthesis}
\label{s:games}

In this section, we formulate quantitative games whose solutions lead to
deterministic implementations of safety controllers that minimize certain cost
functions. Specifically, we introduce {\em mean-payoff games} (MPG)  and {\em
$\lambda$-discounted payoff} games ($\lambda$-DPG) played on the subarena induced by the winning region of the safety game.
In an MPG the goal of the controller is to minimize the limit-average costs accumulated along an infinite play
\begin{equation}\label{e:limaverage}
    \limsup_{T\to\infty} \frac{1}{1+T} \sum_{i=0}^T \gamma(v_i,v_{i+1}),
\end{equation}
and in a $\lambda$-DPG the goal is to reduce the discounted costs
\begin{equation}\label{e:dpg}
(1-\lambda) \sum_{i=0}^{\infty} \lambda^i \gamma(v_i,v_{i+1}),
\end{equation}
where $\lambda\in[0,1)$ is the {\em discount factor}, and $\gamma$ is the cost function on the game.
The intuition is that for $\lambda$ close to $0$ the players only focus on optimizing w.r.t.\ the near future
(as $\lambda^0 = 1$ and $\lambda^k\approx 0$ for $k >> 1$), while for $\lambda$ close to $1$
they focus more on optimizing the cost in the long run.

In the rest of the section we first present these games more formally, and recall some well
known results. Then we show in detail how to construct the games for a given
symbolic system. Moreover, we show how to derive an implementation of a safety
controller from the solution of the 2-player games.
%

\subsection{Mean-payoff games and $\lambda$-discounted payoff games}

Both an MPG and a $\lambda$-DPG are played on an \emph{arena},
which is a weighted directed bipartite graph consisting of the nodes $V=\vmin\uplus\vmax$, the edges $E\subseteq (\vmin\times\vmax\cup \vmax\times\vmin)$ with $\emin:=E\cap \vmin\times\vmax$ and $\emax:=E\cap\vmax\times\vmin$, and $\gamma\colon E\to \Q$.
The nodes $\vmin,\vmax$ belong to the two players $\pmin$ and $\pmax$,
respectively, and each edge $(u,v) \in E$ is assigned a rational cost $\gamma(u,v) \in \Q$ that $\pmin$ has to pay to $\pmax$.
In a play, a pebble is placed on a starting node $v_0$. At each step, the player 
owning the current node chooses a successor of the node, and moves the pebble to it.
A play is an infinite sequence $\{v_i\}_{\N}$ of nodes visited by the moves.
The goal of $\pmin$ is to minimize the (maximal) value of~\eqref{e:limaverage} in an MPG
and the value of~\eqref{e:dpg} in a $\lambda$-DPG, respectively, while
$\pmax$ has the opposite goal.

A \emph{memoryless strategy} for player $\pmin$ is simply a function $\sigma\colon \vmin\to\vmax$ such that $\sigma(v)$ is a successor of $v$ in the arena. Analogously, memoryless strategies for player $\pmax$ are defined.
It is known that for both MPG and $\lambda$-DPG memoryless strategies
suffice to play optimally, i.e.,
there exist memoryless strategies $\sigma_{\min}$, $\sigma_{\max}$
and a valuation $\varv\colon V \rightarrow \R$ s.t.~when $\pmin$ uses $\sigma_{\min}$
to determine to where to move the pebble -- no matter how $\pmax$ chooses to move
-- the resulting average cost for $\pmin$ will be at most $\varv(v)$ for $v$ the node in which the pebble has been placed
initially, and symmetrically for $\pmax$ using $\sigma_{\max}$. It is also known that
for  $\lambda\to 1^-$ the optimal
game values in the $\lambda$-DPG (i.e., the costs of the plays corresponding to the optimal strategies)
converge to the optimal values of the MPG~\cite{FilarVrieze96}.
In fact, there is some $\lambda_0\in [0,1)\cap\mathbb{Q}$
depending only on the given arena such that optimal strategies in the
$\lambda$-DPG w.r.t.\ any $\lambda\ge \lambda_0$ and optimal strategies in the MPG coincide~\cite{Andersson2009}.



\subsection{From valuated safety problems to games}

Let $(\hat S,\hat Z,\hat c, \lambda)$ be a valuated safety problem with
$\hat S=(\hat X,\hat X_0,\hat U,\hat F)$.
As the symbolic model is finite, we can use the well-known fixed point algorithms~\cite{BertsekasRhodes71,Tabuada09}
implemented in {\tt SCOTS}, to solve the abstract safety problem and obtain
a safety controller $\hat K:\hat X\rightrightarrows\hat U$ as a solution of $(\hat S,\hat Z)$.

In order to obtain an implementation of $\hat K$ that optimizes the value for the quantitative problem,
we create an arena (the same for all $\lambda$-DPG including MPG) by associating the controller with $\pmin$,
who chooses the input for a given state, and the environment with $\pmax$, who chooses the successor
in accordance with the transition relation, where the cost of an arena edge is given by the cost function we are using.
As in the next section we consider cost functions that also depend on the last input issued by the controller to the system,
we extend the state space with the last used input\footnote{This in fact means that we consider strategies with bounded memory.}.

With the domain of the controller given by $\domain{\hat K}$ the nodes of the arena are
\begin{subequations}
\label{e:gamearena}
\begin{align}
    \vmin:= \domain{\hat K} \times \hat U \subseteq \hat X \times \hat U,\quad&
    \vmax:= \hat K                        \subseteq \hat X \times \hat U
\end{align}
and edges
\begin{align}
  \emin &:= \{ ((x, u), (x , u')) \in \vmin \times \vmax \mid u' \in \hat K(x) \} \\
  \emax &:= \{ ((x, u), (x', u )) \in \vmax \times \vmin \mid x' \in \hat F(x, u) \}
\end{align}
where the edges in $\emin$ and $\emax$ are assigned the costs
\begin{equation}\label{e:weightfunction}
    \gamma_{\min}((x, u),(x, u')) :=  \hat c(u, x, u') \text{ and } \gamma_{\max}((x,u),(x',u)) := 0,
\end{equation}
\end{subequations}
respectively.

To solve a $\lambda$-DPG with $\lambda \in [0,1)$, we use fixed point iteration on the
fixed point equations derived for both players~(see e.g.\ \cite{FilarVrieze96}).
It follows from Banach's fixed-point theorem that this converges to the least and only
fixed point, and additionally that it converges quickly unless $\lambda$ is close to $1$.
For the case $\lambda = 1$ we solve the resulting MPG 
using the tool presented in~\cite{MeyerLuttenberger16}.

The output in each case is an optimal strategy $\sigma_{\min}: \vmin \rightarrow \vmax$ for $\pmin$.
An implementation of $\hat S$ and $\hat K$ from  $\sigma_{\min}$ is given by
\begin{equation}\label{e:best:controller}
\hat K_{\rm imp}(x,u):=\pi_{\hat U}(\sigma_{\min}(x,u))
\end{equation}
where $\pi_{\hat U}$ is the projection of a pair $(x,u)$ onto $\hat U$, i.e.,
$\pi_{\hat U}(x,u)=u$.

\section{Case Studies}
\label{s:case-studies}

In order to analyse the influence of the discount factor $\lambda$ on the synthesized controller implementation, we study both a HVAC system and the classical cart-pole system in Section~\ref{ss:hvac} and~\ref{ss:cartpole}, respectively.
For both case studies, we will consider the following cost functions:

\begin{align}
    c_{\IS}(u,x,u') &= \begin{cases} 0 & \text{if }u = u' \\  1 & \text{if }u \neq u' \end{cases}
            && \text{(input switches)} \label{e:cost:is}\tag{IS} \\
    c_{\DR}(u,x,u') &= \norm{\pi_r(x) - r}^2
            && \text{(deviation from reference)} \label{e:cost:dr}\tag{DR} \\
    c_{\EC}(u,x,u') &= \norm{u-u_0}^2
            && \text{(energy consumption)} \label{e:cost:ec}\tag{EC} \\
    c_{\ID}(u,x,u') &= \norm{u-u'}^2
            && \text{(input deviation)} \label{e:cost:id}\tag{ID}
\end{align}
These are instantiated with a reference point $r \in \R^m$ (e.g.\ the optimal temperature),
a projection $\pi_r : X \rightarrow \R^m$ onto the coordinates of the reference point and
the input $u_0$ consuming minimal energy (e.g.\ where no actuators are active).

The function $c_{\IS}$ is used to minimize the average amount of input switches and
$c_{\DR}$ is used to minimize the average deviation from a reference point.
These two criteria were already introduced in~\cite{RunggerReissigZamani16}.
The function $c_{\EC}$ assumes that the distance of an input from an energy-minimal input correlates
with the amount of energy consumed by using this input. Then we can use this function to
minimize the average amount of energy consumed. The function $c_{\ID}$ minimizes the change
in successive inputs, which could e.g.~lead to less jerky trajectories. 
We will not analyze the latter two costs seperately, but use them as intermediate functions
to derive the following combined cost function, which is used in~\cite{MeyerGirardWitrant15}:
\begin{align}
    c_{\CC}(u,x,u') &= \frac{1}{3} \left(
            \frac{c_{\DR}(u,x,u')}{\max_{\DR}} +
            \frac{c_{\EC}(u,x,u')}{\max_{\EC}} +
            \frac{c_{\ID}(u,x,u')}{\max_{\ID}} 
        \right) \label{e:cost:cc}\tag{CC}
\end{align}
Here, the normalizing factors
$\max_{\DR}$, $\max_{\EC}$ and $\max_{\ID}$ are the maximal values of the respective cost functions
within the domain of the safety controller $\hat K$.

We derive the cost function $\hat c(u, \hat x, u')$ required by~(\ref{e:cost-abstraction})
by taking the maximum of $c(u, x, u')$ for all $(x,\hat x)\in Q$.
Additionally, as we require rational costs for solving MPGs,
the costs are rounded with a precision of 6 decimal digits.

\subsection{Heat, ventilation and air conditioning}
\label{ss:hvac}

In our first example, we synthesize an implementation of a safety controller for
a heat, ventilation and air conditioning (HVAC) system.
We follow closely the setup described
in~\cite{BrocchiniFalsoneManganiniHolubPrandini16,HolubZamaniAbate16}, which
considers a rooftop
unit that is used to regulate the temperature and to circulate the air in different zones in a
building to keep the air at a comfortable level. The HVAC system
consists of a packaged direct expansion cooling rooftop unit (RTU) that conditions one zone in a single
story building, which is equipped
with a two-stage compressor, a multi-speed fan and modulating economiser
dampers. It is assumed that the economiser dampers remain in constant position and
are not available for control.
A control unit is used to regulate the temperature and the humidity of the
air within the regulated zone within a desired comfort interval, despite the
presence of disturbances.
We refer the interested readers to~\cite{BrocchiniFalsoneManganiniHolubPrandini16,HolubZamaniAbate16} for
a more detailed description of the HVAC.

A linear dynamical system with four states that approximates
the local system behavior at a pre-specified \emph{nominal behavior} 
given by the zone set-points with fixed heating loads, moister loads  and RTU
actions serves as basis of the
design scheme. The nominal dynamics in~\eqref{e:sys:ct} is described by
$f(x,u)=Ax+Bu$, where the  matrices are determined from data during nominal
operation as follows
\begin{IEEEeqnarray*}{l'l}
  A=10^{-4}\cdot
   \left(
   \begin{IEEEeqnarraybox*}[][c]{,r/r/r/r,}
    -28 & -5.6 &   0 &   0\\
      0 & -8.3 &   0 &   0\\
      0 &    0 & -17 &  1 \\
      0 &    0 &   0 & -2.8
   \end{IEEEeqnarraybox*}\right),&
  B=10^{-4}\cdot
   \left(
   \begin{IEEEeqnarraybox*}[][c]{,r/r,}
      -0.8 & -1.7 \\
         0 & 5.8\\
      -1.7 & 0.08\\
         0 & 2.3
   \end{IEEEeqnarraybox*}\right).
\end{IEEEeqnarray*}
The inputs of the system denoted by $\nu_1(t)$ and
$\nu_2(t)$ represent the fan angular velocity and compressor angular velocity, respectively,
and are restricted for all times $t\in\R_{\ge0}$ to the stage values 
\begin{IEEEeqnarray*}{c}
  \nu(t)\in U:=\{-25,0,25,50\}\times\{-50,0,50\}.
\end{IEEEeqnarray*}
In order to account for input uncertainties (which according
to~\cite{HolubZamaniAbate16} are also used to account for model uncertainties) we use
a perturbation bound of $w=B\cdot\left(10,10\right)^\top$ in~\eqref{e:sys:ct}.  Notably our disturbance model is rather
simple yet powerful and unlike
to~\cite{BrocchiniFalsoneManganiniHolubPrandini16,HolubZamaniAbate16}, we do not
assume that the disturbance signal is constant during sampling times.

The first and the third elements of the state vector represent the zone temperature in degree Celsius and the zone relative
humidity in $\%$, respectively. The values are restricted to lie within
$\intcc{-1,1}$ and $\intcc{-5,5}$, respectively. As a result, we obtain as safety specification
\begin{IEEEeqnarray*}{c}
  Z:= \intcc{-1,1}\times\R\times \intcc{-5,5}\times \R.
\end{IEEEeqnarray*}

For the HVAC system, we can interpret the  quantitative specifications as the following:
\begin{enumerate}
    \item \eqref{e:cost:is}: Find a lazy controller that minimizes the amount of input switches.
        In this system, this reduces the wear on the compressor and fan.
    \item \eqref{e:cost:dr}: Find a controller with minimal deviation from the optimal comfort level.
        We set the reference point $r = (0,0)$ and $\pi_r(x) = (x_1,x_3)$, i.e.\ temperature and relative humidity
        should be kept close to the normalized optimal value, which corresponds to 21\textdegree{}C and 50\% humidity.
    \item \eqref{e:cost:cc}: Simultaneously minimize comfort, energy consumption and relative change in inputs.
        The minimal energy input $u_0 = (-25, -50)$ corresponds to having the fan and compressor turned off.
\end{enumerate}

We will construct optimized controllers for each of the cost functions, with different values of $\lambda$, and
evaluate them by comparing the limit-average cost in the long-term when simulating the system.

We introduce a disturbance signal $\omega$ during simulation by instatiating equation~\eqref{e:sys:ct}
by $\dot \xi(t) = f(\xi(t),u) + \omega(t)$, and consider the following disturbance signals:
\begin{IEEEeqnarray*}{rCl}
\dsin(t) &:=& \left(10\sin\left( \tfrac{t}{2\pi\tau} \right), -10\sin\left( \tfrac{t}{2\pi\tau} \right)\right)^T; \\
\dcon(t) &:=& \left(10, -10\right)^T. %
\end{IEEEeqnarray*}
We use $\eta = (0.2, 1, 0.4, 10)$ as the discretization parameter for constructing the symbolic model and
$\tau = 100\,\text{sec}$ as the sampling time.

\subsection{Cart-pole system} 
\label{ss:cartpole}

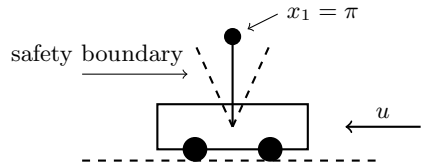
\begin{figure}
\begin{center}
\begin{tikzpicture}[thick]
  \draw (0,0) rectangle (2,.6);
  \draw[fill] (.5,0) circle (.15);
  \draw[fill] (1.5,0) circle (.15);

  \draw[dashed] (-1,-.15) -- (3,-.15);

  \draw[<-] (2.5,.3) -- node[above] {$u$} (3.5,.3);


  \draw  (1,.3) -- (1,1.5);
  \draw[fill] (1,1.5) circle (.1);

  \draw[dashed] (1,.3) -- (1.5,1.4);
  \draw[dashed] (1,.3) -- (.5,1.4);


  \node at (2.2,1.8) {$x_1=\pi$};
  \draw[thin,->] (1.6,1.8) -- (1.2,1.6);

  \draw[->,thin] (-1,1) -- node[above, xshift=-.5cm] {safety boundary} (.4,1);

\end{tikzpicture}
\end{center}
\caption{Cart-pole system. The safety controller ensures that the pole stays
within the interval $\intcc{\tfrac{1}{2}\pi,\tfrac{3}{2}\pi}$.}\label{f:cartpole}
\end{figure}

In this example, we synthesize an implementation of a safety controller that
ensures that the pole, which is attached to a cart, stays within a neighborhood
of the upright position. The
cart-pole system is illustrated in Figure~\ref{f:cartpole}.
We follow~\cite{Reissig10} and describe the system by a four
dimensional differential equation with $f$ in~\eqref{e:sys:ct} given by
\begin{equation*}
\begin{split}
f_1(x,u)&=x_2,\\
f_2(x,u)&=-(\alpha^2\sin(x_1)+u\cos(x_1))-2\beta u,\\
\end{split}\quad
\begin{split}
f_3(x,u)&=x_4,\\
f_4(x,u)&=u,
\end{split}
\end{equation*}
where $\alpha=1$ and $\beta=0.0125$. For this example, we assume that there are
no disturbances and set $\omega=0$. While in~\cite{Reissig10}, 
a reachability problem has been solved to regulate the cart-pole from the
downward facing position to the upright position, we focus on the safety problem
to force the cart-pole to stay in the upright position, i.e., the first state $x_1$ is
constrained to $\intcc{\tfrac{1}{2}\pi,\tfrac{3}{2}\pi}$. As
in~\cite{Reissig10,RunggerStursberg12}, we constrain the third coordinate $x_3$ to
$\intcc{-2.4,2.4}$. For the velocity coordinates $x_2$ and $x_4$ we use the
constraints $\intcc{-1,1}$ and $\intcc{-1.4,1.4}$, respectively. We enforce an input
bound of $U=\intcc{-5,5}$. We use {\tt SCOTS} to synthesize a safety controller
$K$ with the discretization paramater $\eta = (0.05, 0.1, 0.1, 0.1)$ and the
sampling time $\tau = 0.35\,\text{sec}$.
%
%

Here, we can interpret the  quantitative specifications as the following:
\begin{enumerate}
    \item \eqref{e:cost:is}: Find a lazy controller that minimizes the amount of input switches.
        In this system, this avoids frequent changes in acceleration, which causes stress in the system.
    \item \eqref{e:cost:dr}: Find a controller with minimal deviation from the reference point
        $r = (\pi, 0)$ with $\pi_r(x) = (x_1,x_3)$. This reference point corresponds to the pole being
        in upright position and the cart being in the center.
    \item \eqref{e:cost:cc}: Simultaneously minimize deviation from the reference point,
        energy consumption and relative change in acceleration.
        The minimal energy input $u_0 = 0$ corresponds to no acceleration.
\end{enumerate}

As for the HVAC system, we construct optimized controllers for each cost function with different
values of $\lambda$ and compare their long-term performance w.r.t.~these cost functions in the simulation.

\section{Experimental Evaluation}
\label{s:experiments}

For each of the two case studies,
we constructed the symbolic system $\hat S$ and solved the safety game to
obtain a safety controller $\hat K$.
Then for each of the three cost functions $c_{\IS}$, $c_{\DR}$ and $c_{\CC}$
and for different values of $\lambda$,
we solved the resulting $\lambda$-DPG and translated the optimal strategies into an implementation $K_{\rm imp}$.
We choose the values of $\lambda \in \intcc{0,1}$ to include the two extrema:
$\lambda = 0$, which results in a controller greedily optimizing one step, and
$\lambda = 1$, which results in a controller giving an optimal solution for
the limit-average cost on the symbolic system.
Additionally, we chose $\lambda = \nicefrac{1}{2}$, which is the value chosen
in~\cite{MeyerGirardWitrant15}, and several values spaced more closely towards the boundaries.

Then we simulate each controller on its respective system, using two
different disturbance functions for the HVAC system, and measure
the {\em the limit-average} cost for the cost function for which the controller is optimized.
Assessing the performance of the obtained controller w.r.t.\ the limit average has two reasons:
\begin{itemize}
\item 
As the controllers are assumed to run indefinitely on the system,
the limit-average cost is usually the value that is actually the most relevant.
\item 
This allows us to compare controllers obtained for different values of $\lambda$, i.e.\ whether it is preferable to chose $\lambda$ close to $0$ so that $\pmin$ only optimizes w.r.t.\ the near future or $\lambda$ close or equal to $1$ so that the far future becomes more and more important for $\pmin$.
\end{itemize}
In the simulations, the systems display periodic behaviour, and
we run the simulation sufficiently long enough until the limit-average cost stabilizes.

The obtained values are summarized in Table~\ref{tb:results}.
The sizes of the symbolic models and times needed to construct them, to construct the arena, and to solve the respective games are given in Table~\ref{tb:size-times}.
\begin{table}[ht]
\begin{center}
\caption{Size of the symbolic model for each system and
times in seconds to construct the safety controller $\hat K$, the game arena $G$, and the maximum times over all cost functions $c$
to solve the $\lambda$-DPG for any $\lambda \le \nicefrac{15}{16}$ and to solve the MPG.}
\label{tb:size-times}
\scalebox{0.90}{\begin{tabular}{p{1.5cm}*{3}{r@{\hskip 0.8cm}}*{4}{r@{\hskip 0.8cm}}}
\toprule
& \multicolumn{3}{c}{Size of $\hat S$} & \multicolumn{4}{c}{Time (sec)} \\
\cmidrule(r){2-4} \cmidrule(r){5-8}
System     & $|\hat X|$ & $|\hat U|$ & $|\hat F|$ & ${\hat K}$ & $G$ & $\lambda$-DPG & MPG \\
\midrule
HVAC       & $1.4\cdot10^3$ &  12 & $3.8\cdot10^6$ & 16.78 & 55.37 & 8.49 &  477 \\
Cart-pole  & $2.3\cdot10^6$ & 101 & $3.1\cdot10^9$ &  3558 &  1306 &  156 & 8846 \\
\bottomrule
\end{tabular}}
\end{center}
\end{table}

\begin{table}[ht]
\begin{center}
\caption{Values of the limit-average cost for different cost functions and controllers on the two systems.
    For each system and cost function $c$, we synthesize controllers for each given value of $\lambda$.
    We then simulate the controller, possibly with different disturbance signals $\omega$, and measure
    the limit-average cost w.r.t.~the cost function $c$.
    The entry for each value of $\lambda$ lists the result of this limit-average cost with the
    respective controller. The entry for $\varv$ lists the upper guaranteed bound on the limit-average cost by the controller from
    the MPG with $\lambda = 1$. The best values in each row are marked in bold.}
\label{tb:results}
\scalebox{0.90}{\begin{tabular}{p{1.5cm}ll*{10}{r@{\hskip 0.19cm}}}
\toprule
& & & \multicolumn{9}{c}{$\lambda$} & \\
\cmidrule(r){4-12}
System & 
$c$ & $\omega$ &
$0$ & $\nicefrac{1}{16}$ & $\nicefrac{1}{8}$ & $\nicefrac{1}{4}$ & $\nicefrac{1}{2}$ & $\nicefrac{3}{4}$ & $\nicefrac{7}{8}$ & $\nicefrac{15}{16}$ & $1$ & $\varv$ \\
\midrule
\multirow{6}{\linewidth}{$\text{HVAC}$}
& \multirow{2}{*}{$c_{\IS}$}
& $\dsin$ & 0.082 & \textbf{0.051} & \textbf{0.051} & \textbf{0.051} & \textbf{0.051} & \textbf{0.051} & \textbf{0.051} & \textbf{0.051} & \textbf{0.051} & \multirow{2}{*}{\em 0.750\/} \\
& & $\dcon$ & 0.278 & \textbf{0.167} & \textbf{0.167} & \textbf{0.167} & \textbf{0.167} & \textbf{0.167} & \textbf{0.167} & \textbf{0.167} &         0.175 &   \\
\cmidrule(r){2-13}
& \multirow{2}{*}{$c_{\DR}$}
& $\dsin$ & 2.631 & \textbf{0.069} & \textbf{0.069} & \textbf{0.069} & \textbf{0.069} & \textbf{0.069} & 0.079 & 0.165 & 1.919 & \multirow{2}{*}{\em 5.410\/} \\
& & $\dcon$ & 12.784 & \textbf{0.109} & \textbf{0.109} & \textbf{0.109} & \textbf{0.109} & \textbf{0.109} & 0.126 & 0.137 & 2.757 &  \\
\cmidrule(r){2-13}
& \multirow{2}{*}{$c_{\CC}$}
  & $\dsin$ & 0.147 & 0.137 & 0.136 & 0.132 & 0.130 & 0.126 & \textbf{0.119} & 0.123 & 0.148 & \multirow{2}{*}{\em 0.796\/} \\
& & $\dcon$ & 0.346 & 0.346 & 0.346 & 0.346 & 0.336 & 0.261 & \textbf{0.170} & \textbf{0.170} & 0.213 & \\
\midrule
\multirow{3}{\linewidth}{Cart-pole}
& $c_{\IS}$ & & 0.765 & 0.380 & 0.375 & \textbf{0.373} & \textbf{0.373} & \textbf{0.373} & 0.377 & 0.389 & 0.765 & {\em 1.000\/} \\
\cmidrule(r){2-13}
& $c_{\DR}$ & & 1.610 & 1.378 & 1.377 & 1.314 & 1.113 & 0.031 & 0.030 & \textbf{0.023} & 1.246 & {\em 5.117\/} \\
\cmidrule(r){2-13}
& $c_{\CC}$ & & 0.046 & 0.045 & 0.044 & 0.032 & 0.032 & \textbf{0.005} & 0.009 & 0.013 & 0.034 & {\em 0.625\/} \\
\bottomrule
\end{tabular}}
\end{center}
\end{table}

In our experiments, 
the controller with an optimal choice of $\lambda$ is sometimes up
to 50 times better than a controller for a different $\lambda$.
Also, there is no single optimal choice
for $\lambda$: depending on the system (resp.\ the symbolic model) and the cost function, it may
be necessary to choose different values.
In particular, even though the controller for $\lambda=1$ gives an optimal
solution for the \emph{symbolic system}, its actual performance
on the \emph{continuous-time system} is almost never optimal when compared to
the implementations obtained for smaller values of~$\lambda$.

Note that our approach can handle systems with up to several million states and
a billion transitions (see Table~\ref{tb:size-times}), and that most of the computation time is spent on constructing
the safety controller and the unweighted arena (i.e.\ without edge costs). As this basic arena is independent
of the choice of $c$ and $\lambda$, it can be reused to solve different $\lambda$-DPGs.
Solving such a game is fast unless $\lambda$ tends to 1, and therefore our tool chain 
allows to easily construct and evaluate controllers for several different values of $\lambda$
and choices of the cost function $c$.

Our explanation for the subpar performance of the controller obtained for $\lambda=1$ 
is that this reflects the conservative over-approximation used in the construction of the symbolic model.
In the game on the symbolic system, the adversary player ($\pmax$)
chooses successors for sequences of inputs from the symbolic states,
which may lead to transition sequences that can never occur in the
concrete system.
This overestimation of reachable states increases as the length
of the sequence increases.
In the case of an MPG, the controller ($\pmin$) optimizes its inputs
to guard against the worst-case for arbitrarily long sequences.
It may choose any strategy that achieves the optimal value $\varv$,
even a very conservative one. As seen in Table~\ref{tb:results}, the
optimal value of the MPG is usually much higher than the actually achievable
minimal costs. To support our hypothesis, we tried to simulate the worst-case
behaviour by extracting the optimal strategy $\sigma_{\max}$ of $\pmax$ from
the MPG for the HVAC system and testing if we could chose a disturbance
$\omega(t) \in \left( \intcc{-10,10} \times \intcc{-10,10} \right) \cap \left( \Z \times \Z \right)$
in each time step leading to a successor in accordance with $\sigma_{\max}$.
However, this was impossible most of the time, and we could not enforce the worst-case costs.
While this approach does not cover all possible disturbance signals, it still
gives an indication that many transitions of the symbolic system can not occur
in the concrete system.

To reduce this gap, one might choose a finer
approximation, however this is often impossible due to the increasing size of the symbolic system.
Using discounted payoff functions comes at no additional cost and leads
to better performing controllers. This is due to their focus on optimization towards
the near future, which offsets the error from the overapproximation.





\section{Summary}
\label{s:summary}

In this paper we studied different quantitative implementation strategies of
safety controllers for perturbed, nonlinear, control systems.
We use normalized $\lambda$-discounted costs in order to define the costs accumulated along a run
of the system. The normalization allows us to also cover the limit-average costs
for $\lambda\to1^{-}$ thereby also obtaining a unified presentation of previous
results. 
We present two case studies and conduct a systematic analysis to
illustrate the effects of different choices of cost functions and
discount parameters.
We show that carefully choosing $\lambda$ allows us to reduce the
limit-average cost associated with the refined controller 
quite drastically when compared to the fixed values of
$\lambda=\frac12$ and $\lambda\to 1^{-}$ found
in~\cite{MeyerGirardWitrant15} and \cite{RunggerReissigZamani16}, respectively. Our explanation for
this is that by carefully choosing $\lambda$ the safe, yet pessimistic
over-approximation used for the construction of the symbolic model can be offset at
least to some extent. With our existing tool-chain it is effectively possible
to sample for different choices of $\lambda$ in a reasonable amount of time for
symbolic models consisting of up to several millions of states.

\bibliographystyle{splncs03}
\bibliography{references}

\end{document}